\documentclass[12pt,letterpaper]{article} % Vancouver Reference Style (Numbered)
\usepackage{url,hyperref,lineno,microtype,caption,subcaption,graphicx}
\usepackage[onehalfspacing]{setspace}

\date{}

\author{J. Alimena\,$^{1}$, \and C. Beyer\,$^{1}$, \and B. Br{\"u}ers\,$^{1}$, \and F. Gaede\,$^{1}$, \and N. Gillwald\,$^{1}$, \and M. Hernandez Villanueva\,$^{2}$, \and E. Jones\,$^{1,*}$, \and Y. Kemp\,$^{1}$, \and T. Madlener\,$^{1}$, \and K. Schwarz\,$^{1}$ \and and C.~Wissing\,$^{1}$ \\ 
\small $^{1}$ Deutsches Elektronen-Synchrotron DESY, Notkestr. 85, 22607 Hamburg, Germany \\ 
\small $^{2}$Brookhaven National Laboratory, Upton, New York 11973, U.S.A. \\\\
\small $^{*}$Corresponding author: Eleanor Jones; eleanor.jones@desy.de
}

\title{Sustainable computing workshops in high-energy physics at DESY}

\begin{document}
\onecolumn

\maketitle

\begin{abstract}
This perspective article details the program of sustainable computing workshops launched within the High Energy Physics department at Deutsches Elektronen-Synchrotron (DESY) in 2023. The workshop series targets scientific users of the National Analysis Facility, hosted at DESY, in order to promote sustainable software development and usage across a wide range of applications. Details of the structure of the workshops, as well as the reception amongst participants, will be presented. In addition, plans for the expansion of the workshop series will be outlined.
\end{abstract}

\clearpage
\tableofcontents
\clearpage

\section{Introduction}\label{sec:intro}
The Deutsches Elektronen-Synchrotron (DESY) is one of the largest research centres in Germany, where a wide range of research is performed. Amongst the diverse topics, high-energy physics (HEP) is a significant part of the program, and this has large computing and storage demands. Since 2007, DESY has hosted the National Analysis Facility (NAF), designed to analyse and simulate data from the particle physics experiments ATLAS~\cite{ATLAS:2008xda}, CMS~\cite{CMS:2008xjf}, Belle~II~\cite{Belle-II:2010dht}, LHCb~\cite{Lhcb_2008}, past and present on-site experiments, and from possible future colliders such as the International Linear Collider and the Future Circular Collider. The NAF has over 1000 registered scientific users of the large batch system, thus complementing the DESY Grid by allowing interactive and fast analysis of large data sets on Grid storage systems.

The ongoing and immediate climate crisis necessitates a drastic reduction of the carbon footprint in all areas of society, a statement that many governments have acknowledged through the signing of the Paris Agreement~\cite{unfccc2015paris}. It is clear that human activities, in particular through emissions of greenhouse gases, are the cause of global warming~\cite{ipcc2023ar6}. Therefore, it is both a collective responsibility and an individual responsibility to reduce this impact. 

Within HEP, one of the areas that can be targeted to reduce the carbon footprint of the field is computing~\cite{Banerjee:2023avd}. It was estimated that the information and communication technology sector (ICT) was responsible for 2--6\% of global CO$_2$ emissions in 2020, which could grow to 20\% by 2030~\cite{KERN201553,ADISA2024141768,BELKHIR2018448,Zulfiqar2023}. Most of the efforts to reduce the carbon footprint of ICT have been dedicated to improving hardware energy efficiency. However, it is largely agreed that improving the sustainability of ICT will also need to involve making software more sustainable by reducing the energy consumption or the carbon footprint associated with its deployment~\cite{Danushi2024}. The potential impact of ``Green IT'' in general, covering hardware and software, and ``Green Coding'', covering ecologically sustainable software in particular, is explained in several review articles such as Refs.~\cite{Danushi2024,Junger2024}.

In 2023, an initiative was launched within the HEP group at DESY to provide a series of workshops targeting users of the NAF. The workshops aim to educate scientific users on software and batch computing best practices, in order to encourage software development across a wide range of applications with the aspect of sustainability in mind. While developing research software requires different skills than operating the software on large scale computing systems, an optimal interplay between these two areas is necessary for both positive user experience and positive environmental impact.

The term ``sustainability'' was defined in 1987 by the United Nations Bruntland Commission~\cite{brundtland1987our} as ``meeting the needs of the present without compromising the ability of future generations to meet their own need''. It is an umbrella term that is often used to refer to the many diverse actions, in science and society, that can be taken to address the climate crisis and its impacts. In this series of workshops at DESY and throughout this paper, we use the terms ``sustainability'' and ``sustainable computing'' mainly to refer to two related aspects; on the one hand, this refers to environmental sustainability, on the other hand, we also use it to cover aspects of the software development life cycle. The former mainly targets power conservation, intelligent software development, and computing in an environmentally-sustainable way, while the latter aims to instil the basics for writing robust and long-term maintainable software. The efforts towards environmental sustainability can be directly mapped to the UN's sustainability Goals~\cite{un2015sdgs}, most importantly Goal 12 and Goal 13, i.e. to ``ensure sustainable consumption and production patterns'' and to ``take urgent action to combat climate change and its impacts'', respectively. Software sustainability aspects cannot as easily be mapped to environmental aspects, but are indeed part of a systemic approach to sustainability~\cite{karlskrona_paper}\footnote{See also \url{https://sustainabilitydesign.org/}}. A few efforts, such as attempts to show the effects of code on energy consumption~\cite{Mancebo2021,Verdecchia2018,Junger2024}, have started to quantify the impact of software sustainability on environmental sustainability. However, concrete practical guidelines for sustainable software are not yet commonly available and are still part of active research~\cite{Danushi2024,Junger2024}.

Initially, the series was launched with a beginners' workshop dedicated to providing a basic understanding of how to write code and how to perform batch computing in a more sustainable way. This workshop targets new users of the NAF and focuses predominantly on common tools and software utilised within HEP and their use within the NAF. 

Following the success of the initial workshop, the series has continued to provide the beginners' workshops, as well as providing dedicated advanced workshops in specific areas of software development practices. The advanced workshops aim to provide methods to optimise code performance and to reduce the impact of software bugs, with, for example, software testing.

This perspective article will provide an overview of the sustainable computing workshop series launched within HEP at DESY in 2023, focusing on the plans for expansion and progression of the series, including bringing the program outside of DESY. 

\section{Computing on the National Analysis Facility}\label{sec:NAF}
DESY offers large computing resources for the communities it hosts. The Interdisciplinary Data and Analysis Facility (IDAF) is the umbrella under which several infrastructures are set up to exactly target the needs of the communities. The NAF was set up in 2007 in the realm of the Helmholtz Alliance ``Physics at the Terascale'', and serves the German ATLAS and CMS communities, the global Belle~II collaboration, as well as a variety of small- and medium-sized, DESY-based experiments. The NAF is optimised for interactive and fast turn-around analysis of large data sets. It benefits from the proximity of another IDAF component, serving as a `Tier-2' site for the Worldwide Large Hadron Collider (LHC) Computing Grid (WLCG)~\cite{WLCG} for ATLAS, CMS, and LHCb, as well as a RAW data centre for Belle~II. Data and data access are central to any HEP analysis, and having common data for the NAF and the WLCG serves users from both systems.

The NAF itself is composed of the following triad:
\begin{itemize}
    \item flexible login and access systems (small sized);
    \item batch compute system (large sized);
    \item scaling and flexible storage system, integrated into and shared with experiments' data management.
\end{itemize}
The current access services to NAF are SSH servers, remote desktop and screen sharing, and Jupyter hub. Further services can be added if needed. Dedicated resources need to be reserved, which might lead to inefficient usage in the case where reserved resources remain idle. We therefore integrated the interactive Jupyter service into the batch farm~\cite{jupyter-batch}.

The batch computing system is at the heart of the NAF. A large pool of identically configured compute servers is governed by a scheduling system, that is set up such that it honours group fair share, and favours short, interactive-like jobs over large production jobs. For versatility purposes, the batch farm also supports jobs running in containers provided by users or experiments.

Last but not least is the storage system, which on the NAF consists of several parts. At the centre is the ``dCache'' system~\cite{dcache}, which holds the large data that is central to both the WLCG and the NAF. Secondly, there is the `DUST' system, which is local to the NAF and contains project space for users and groups. The AFS storage system~\cite{AFS} provides user access to a global file system, whereas CVMFS~\cite{CVMFS} is a global file system for centrally-organised software and container image distribution. For ease of usage, all of these storage systems are presented as mounted file systems with all features of the Portable Operating System Interface (POSIX) access model. Storage system usage is one prime example of the interplay between software development and cluster usage. Whilst during the development process, for example on a laptop, only the functionality of the software is checked for performance, when considering the cluster usage, the storage access on a large scale also needs to address performance considerations. As an illustration of this principle, the number of open/close operations on a file should stay minimal for a network file system because of the higher access latency compared to a local solid-state drive, for example. 

The NAF is operated by the DESY IT department. Amongst their duties, and a key factor of the success of the NAF, are extensive support, documentation, and consulting. Furthermore, other factors are also important for the optimal use of the NAF, such as good code development workflows, usage of optimal algorithms and programming techniques, and well-tested code. User support and training on these aspects is best provided by the experiments and expert users --- optimally in a way that it complements the NAF infrastructure. Until the introduction of the workshops described in this paper, a formal training platform had not been available at DESY.

Raising awareness that computing has a substantial environmental impact is important~\cite{TenRules}, and one effort that the IT department at DESY has initiated is to provide users of the NAF with an estimate of their CO$_2$ footprint produced by their computational tasks, in the form of a weekly summary. This initiative has been well received and will be refined in the future to also take storage access into account.

The RF2.0 project~\cite{rf20} aims to design, develop, and validate novel solutions for components and systems, allowing for particle accelerators to be transformed into more sustainable and energy-efficient research infrastructures. In the context of this project, the DESY IT department will investigate the dynamic provisioning of resources and adaptation to both the user load and green energy availability. Future sustainability workshops might involve training on how to best prepare jobs for such a dynamic provisioning.

\section{Overview of the workshops}\label{sec:workshopsOverview}
As highlighted above, sustainable computing necessitates user training on best practices, not only in programming but also in optimal use of resources within a computing system such as the NAF. The goal of the workshops provided within HEP at DESY is to promote sustainable practices within personal use of the NAF. To do so, the workshops present the concepts in short talks, followed by hands-on exercises. These talks and exercises are reused and improved with each iteration of the workshop, providing a persistent resource that can be used even outside the workshop context. The workshop series has so far consisted of three beginners' workshops and one advanced workshop, both of which will be explored in further detail below.

\subsection{Beginners' workshops}\label{sec:beginnerWorkshops}
The beginners' workshop is targeted primarily at students and early-career post-doctoral researchers who are new to the NAF environment. Attendees get an introduction to some of the commonly used software tools within HEP and information that is specific to working in the NAF environment with sustainability in mind. Several of the sections build on the experience gained in earlier sections of the workshop. There is a strong emphasis on building a coherent path through the different sections to create links between all the different elements and how they all contribute to reducing the environmental impact of the NAF. 

\subsubsection{Introduction to ROOT}\label{sec:root}
ROOT~\cite{ROOT} is a software framework that is an integral part of data analysis in HEP, allowing fast access to huge amounts of data. Within HEP, two of the most common ways to use ROOT are interactively through the command line and integrated into existing languages such as C++ and Python. 

Within the beginners' workshop, the attendees are given a broad overview of the use of ROOT, through several interactive exercises to demonstrate the capabilities of the language on the command line. This includes introductions to the types of objects that can be stored in a ROOT file, how to display their information, and how to perform basic manipulations of the stored data. Furthermore, attendees are also guided through creating a basic ROOT macro that allows for fast data visualisation, as well as being provided an introduction to PyRoot through Jupyter notebooks.  

\subsubsection{Batch computing on the NAF}\label{sec:batchComputing}
Batch systems are very powerful because they grant access to a large amount of central processing unit (CPU) resources. However, when used inappropriately, there is the risk of huge resource waste or even spoiling the whole infrastructure~\cite{TenRules}. The workshop covers an introduction to HTCondor~\cite{condor-practice}, the batch system used for the NAF, with some ``dos and don'ts''. 

During a practical hands-on exercise, attendees process a few terabytes of a CMS OpenData~\cite{OpenData-Dimoun} sample to convert the input data from the CMS AOD~\footnote{AOD stands for Analysis Object Data and it is a data format that contains reconstructed `physics objects' with all the information required for a CMS analysis.} format to a small ROOT tree. These concepts were introduced in the ROOT part of the workshop, and used again in this exercise on batch computing. For the exercise, an example from the CERN OpenData portal was adopted for use on the NAF, using data collected by the CMS experiment from 2009 to 2012 (LHC \mbox{Run 1}). Since the CMS software release required to process the Run~1 data cannot run on recent and supported Linux distributions, the exercise also involves how to employ Apptainer~\cite{Apptainer} containers on the NAF batch system, as a practical example of persistent data analysis.

\subsubsection{Introduction to Git}\label{sec:git}
Git~\cite{git} is one of the most used version control systems, which allows users to keep the history of their code and collaborate easily with others. Especially in large, distributed developer teams as found in HEP, such an approach is important~\cite{CollaborativeDevelopment}. Git is ``distributed'', in that each developer has a copy of the repository, the ``local'' repository, where they make changes, and then they interact with one or more ``remote'' repositories to share their changes or import changes from others. Version control systems like Git allow users to collaborate efficiently on software developments, providing a reliable platform for future developments and code preservation.

During the workshops, participants listen to a brief lecture on Git that illustrates the motivation for using such version-controlled documentation for software development and then introduces basic Git commands. Several ways of organising work in Git are demonstrated. Then, the workshop participants attempt a hands-on tutorial to set up and begin to use Git. In this tutorial, the participants learn how to clone, fork, branch, merge, and rebase an example code repository.

\subsubsection{Best coding practices}\label{sec:codingPractices}
Software in HEP is usually written in two languages: C++ is used for anything performance critical and is usually wrapped with Python to allow for a more convenient configuration. Given that the majority of students do not receive any formal education in software development during their physics studies apart from some introductory courses, most of them pick up their skills along the way. A common experience is to get handed an existing script or program and attempt to integrate some new task. Very often these scripts have grown somewhat organically without any regard for maintainability, as long as ``it gets the job done''. A considerable amount of precious person-power is lost this way because changes require manual tests and validation without any real guarantees that existing functionality is not broken. Additionally, computing resources are easily wasted because these codes are often hard to configure and use correctly, resulting in unnecessary test runs or even large-scale batch job failures due to undetected software bugs, the environmental impact of which has been shown in Ref.~\cite{Fister2024}.

To mitigate the main pain points in this common workflow, the major focus of these exercises is to instil some generally applicable, mostly language-agnostic good practices, for example, how to structure software such that it becomes (automatically) testable and maintainable. Additionally, we introduce the basics of modern CPUs and the implications of writing performant software by considering memory layout and considerations for running on multiple threads. Finally, we highlight some common performance pitfalls when writing code in C++ and how to avoid them.

\subsubsection{Continuous integration}\label{sec:CI}
Continuous integration (CI) is a software development practice in which automated builds and tests are performed every time a contributor merges their changes into a remote repository using version-control programs like Git. These tests typically ensure that the code continues to work by performing specific tests such as checking that C++ code compiles, unit tests of limited scope, and so on. CI provides the mechanism for these types of tests to be executed automatically. One relevant example of how CI can be used to support sustainable computing in HEP is to test analysis jobs before submitting large-scale jobs to a batch system, which would avoid wasting resources on jobs that fail or produce meaningless output. CI could also be used to ensure an analysis is reproducible by testing that the output is the same when a new version of a dependent package is deployed or when the code runs on a different machine or computing cluster.

In the beginners' workshop, the attendees are introduced to the basic concepts behind CI through a brief lecture followed by hands-on exercises based on the training exercises developed by the HEP Software Foundation (HSF)~\cite{Malik:chep2023,Malik:2021lgv,HEPSoftwareFoundation:2018unb}. The exercises focus on GitLab~\footnote{GitLab is a platform offering Git repositories and collaboration features. Other platforms also exist; however, GitLab is a common platform within the HEP experiments and at DESY.} CI as a concrete example. The attendees learn how to write a simple configuration file in Yet Another Markup Language (YAML)~\cite{yaml}, which is a human-readable, data-serialization language. Once such a configuration file is available, it can be used as input to a job, which defines what to run in a pipeline, which is the top-level component of CI, delivery, and deployment. Once the attendees learn how to execute a simple configuration file in a pipeline, layers of complexity are added, demonstrating how to build and execute code with CI, use a container image, build multiple versions of the code, and build only when changes are introduced. More complicated concepts like stages and artefacts are introduced for those who want further enrichment.

\subsection{Advanced workshops} \label{sec:advancedWorkshops}
The advanced workshops focus on one topic and treat that topic in more depth than the beginners' workshops. To allow for participation from a wider audience, including staff scientists and senior post-doctoral researchers, these workshops last only a half day. To date, one advanced workshop has been presented on software testing, detailed below. The workshop series is aiming to expand to also include advanced workshops on focused software engineering topics, depending on interest from the HEP community at DESY.

\subsubsection{Software testing}\label{sec:softwareTesting}
This workshop is targeted at an audience that is already familiar with the basics of writing software in either Python or C++ and that is interested in how software can be tested. The main goal of this advanced workshop is to familiarise participants with the terminology of software testing and to make them aware of its benefits. Apart from the obvious benefit of a reduced chance of introducing bugs into the code, the increased maintainability, easier transitions of responsibility for a piece of software, and also the fact that tests are very effective in easing the debugging process are highlighted. The workshop is also used to show participants good practices in setting up software projects and how to organise them, such that they become more easily usable by others as well. This allows for a distribution of developers on the same project, in a way that is fully established by large software projects inside and outside of HEP.

After the brief theoretical overview, the rest of the workshop focuses on a practical introduction to writing unit tests and integrating a test suite into a CI pipeline (discussed in Section~\ref{sec:CI}). For this, several exercises with increasing complexity were prepared, ranging from a very basic introduction to the pytest~\cite{pytest-doc} and Catch2~\cite{catch2-github} unit testing frameworks, to advanced topics like instrumenting tests with \emph{sanitizers}\footnote{Sanitizers are software tools that are used for discovering bugs or suspicious behaviour at runtime by inserting instrumentation code or checks during compilation.} to detect issues that can go unnoticed even if unit tests are passing. This way, this half-day workshop is able to cover various levels of prior knowledge that participants have.

\section{Analysis of the workshops}\label{sec:workshopsAnalysis}
The workshop concept of combining short introductory talks (providing a basic level of knowledge and context to all participants) with a concise, well-documented set of hands-on exercises and solutions (which can be continued after the workshop and used as future reference material) has been very successful. This concept has also enabled a smooth transition between different team members teaching subtopics of the workshop, which is especially important in an environment, such as academia, with rapid changes in the organising team. The split into general beginners' workshops and dedicated advanced workshops allows better coverage of interesting topics, although stand-alone deep-dives into a specific topic can pose difficulties when there is no time to explain other concepts to a certain level.

Participant feedback was gathered using surveys shared with the participants before and after the workshops. All post-workshop surveys included free-text fields to allow participants to share their general perception of the workshop, the positive aspects, and the possible shortcomings. The beginners' workshop survey included additional questions on the self-perceived proficiencies of the participants ranked on a scale from one to five. They were asked before and after the workshop so that the level of improvement could be inferred. Answers were only included in the proficiency analysis if participants had answered the question both times. The advanced workshop survey included additional questions related to the amount of learning, and the confidence in applying the learned concepts, also ranked on a scale from one to five. For all questions requiring a score from one to five, there were no labels given to the scaling other than the extremal values in either survey; therefore, the specific meaning of the values of two, three, and four is subject to interpretation by the survey participants.

\subsection{User opinions}
The workshops were well received in general. Both the participants of the beginners' and the advanced workshops liked the extensive use of easy-to-follow, well-documented, and self-explanatory exercises, providing detailed explanations and including example solutions. Practical tips from and discussions with experts were greatly appreciated, and participants anticipated an easy transfer of skills learned in the workshop to their own future work.
\begin{figure}
    \centering
    \subfloat[C++ proficiency.]{
        \includegraphics[width=0.45\textwidth]{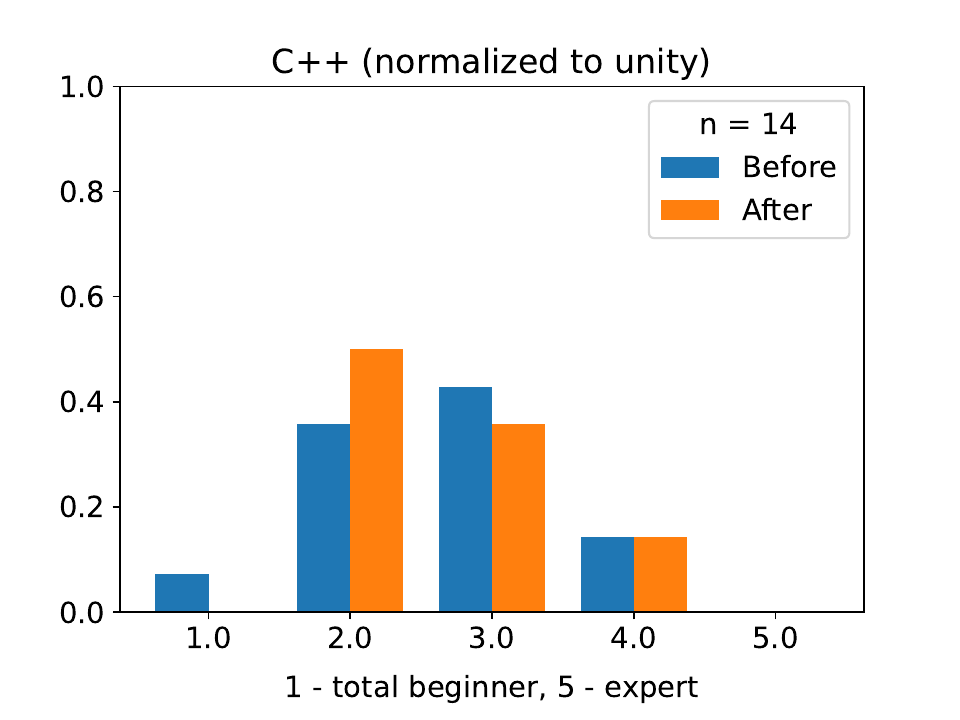}
        \label{fig:analysis:experience_surveys:cpp_matched_normed}
    }
    \subfloat[Python proficiency.]{
        \includegraphics[width=0.45\textwidth]{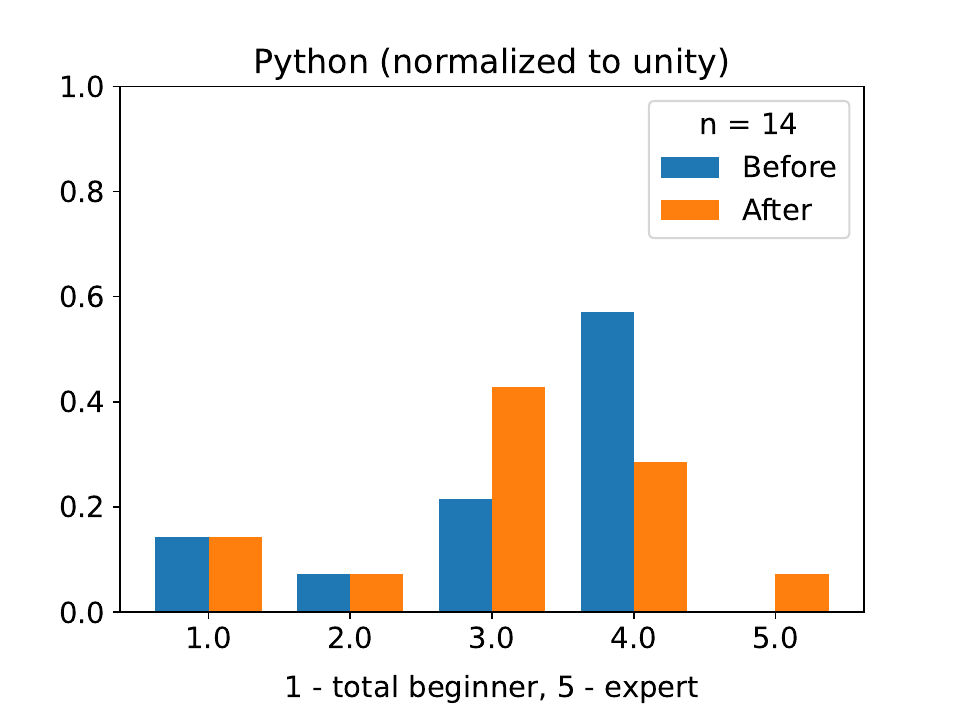}
        \label{fig:analysis:experience_surveys:python_matched_normed}
    } \\
    \subfloat[Git proficiency.]{
        \includegraphics[width=0.45\textwidth]{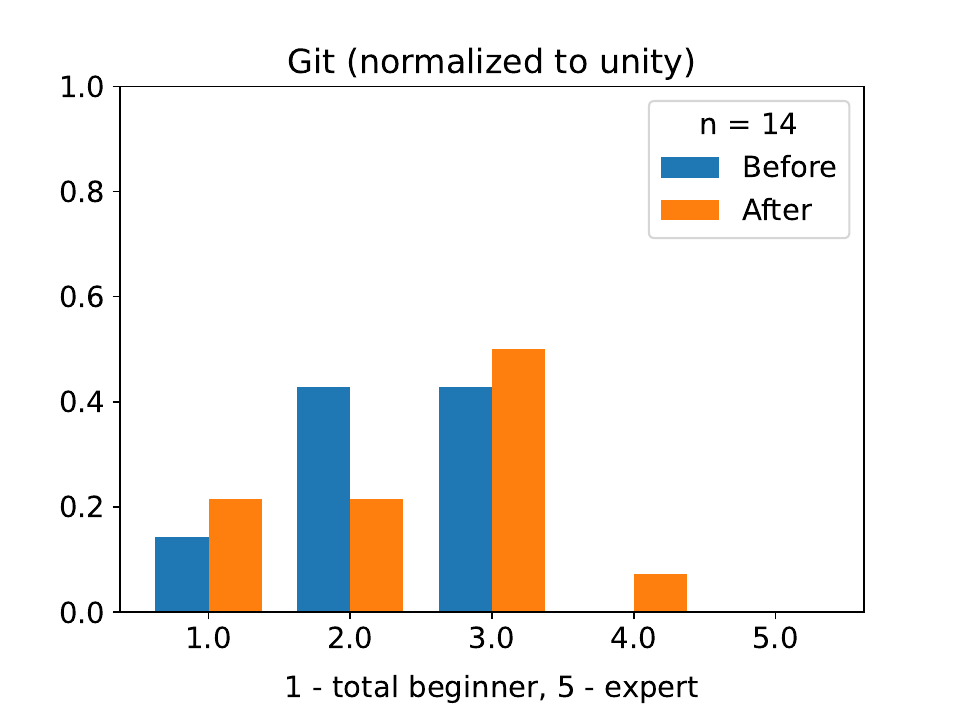}
        \label{fig:analysis:experience_surveys:git_matched_normed}
    }
    \subfloat[HTCondor proficiency.]{
        \includegraphics[width=0.45\textwidth]{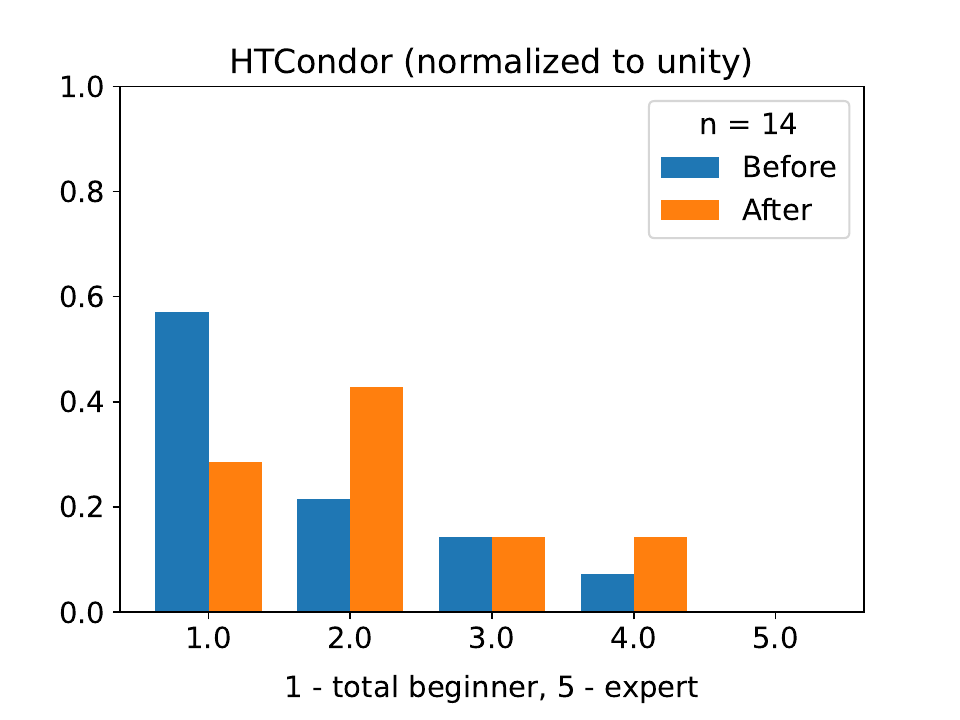}
        \label{fig:analysis:experience_surveys:condor_matched_normed}
    } \\
    \subfloat[ROOT proficiency.]{
        \includegraphics[width=0.45\textwidth]{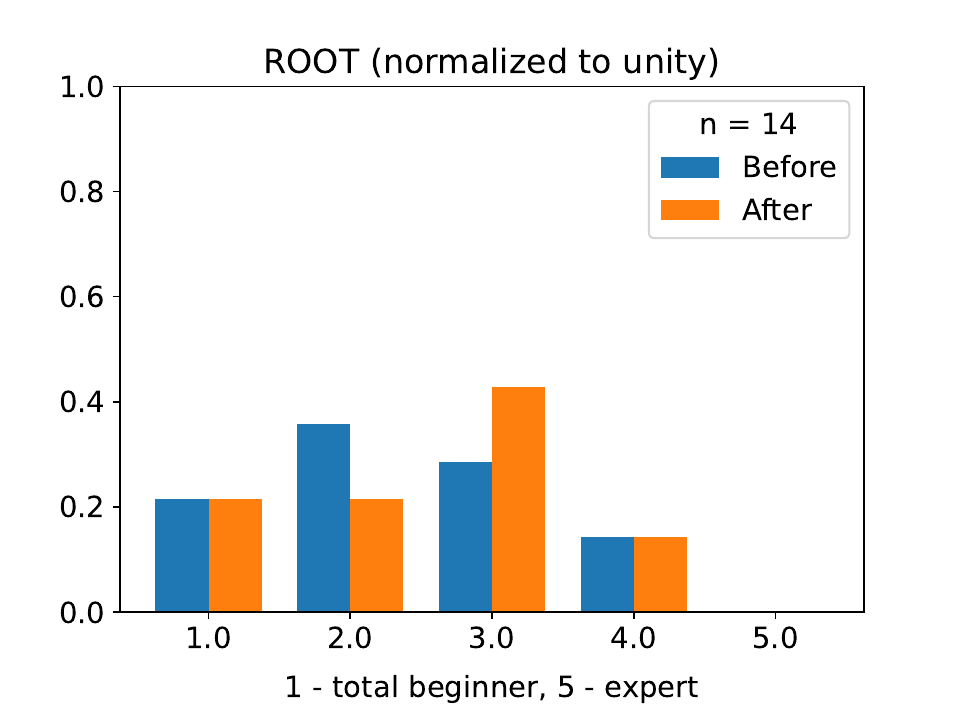}
        \label{fig:analysis:experience_surveys:root_matched_normed}
    }
    \caption{Survey on the participants' self-perceived proficiency in different skills taught at the second instalment of the beginners' workshop in January 2024. The question: ``On a scale of 1 (total beginner) to 5 (expert), how would you assess your competency with \dots?'' was asked. The data includes only participants who filled out the survey both before and after the workshop.}
    \label{fig:analysis:beginner_surveys}
\end{figure}
\begin{figure}
    \centering
    \subfloat[``On a scale of 1 (nothing) to 5 (a lot), how much more do you feel you know about unit testing after participating in the workshop?'']{
        \includegraphics[width=0.45\textwidth]{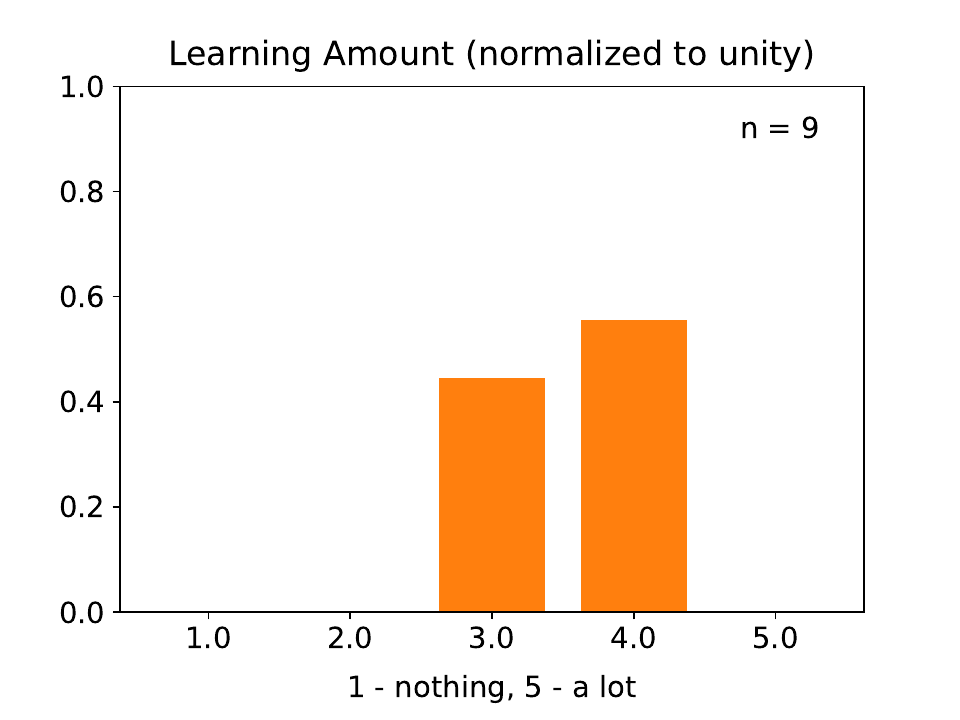}
        \label{fig:analysis:advanced_surveys_a}
    }
    \hfill
    \subfloat[``On a scale of 1 (unsure) to 5 (confident), how would you assess your competency to now write unit tests for your own code base?'']{
        \includegraphics[width=0.45\textwidth]{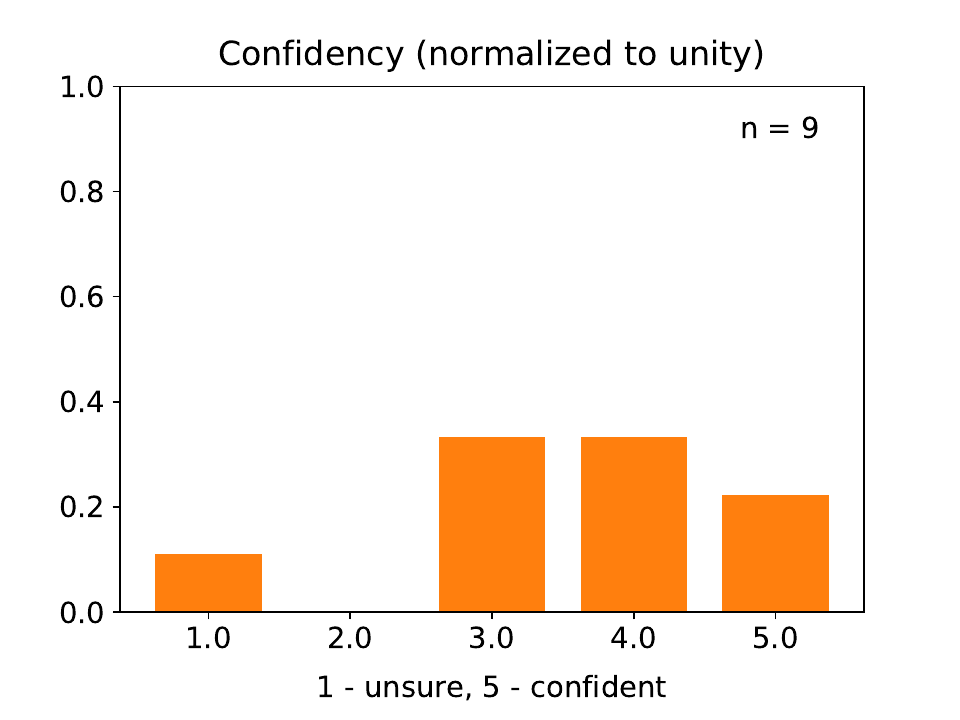}
        \label{fig:analysis:advanced_surveys_b}
    }
    \caption{Survey on the participants' self-perceived knowledge about unit tests and their confidence to implement them in their own future work after having attended the advanced workshop on unit testing.}
    \label{fig:analysis:advanced_surveys}
\end{figure}

Participants of the second beginners' workshop rated their competency with HTCondor, Git, and ROOT higher after the workshop than before, while there was no perceived improvement in C++ and Python knowledge, as shown in  Figure~\ref{fig:analysis:beginner_surveys}. Participants of the advanced workshop felt more confident about testing after the workshop, and felt confident to write tests for their own code base, shown in Figure~\ref{fig:analysis:advanced_surveys}. The half-day length of the advanced workshop was complimented as making it easier to find time to attend the workshop.

Possible improvements to the beginners' workshops mostly relate to the participants' heterogeneity in knowledge and thus the perceived level of the workshop. Therefore, it might be useful to provide more detailed initial exercises for total beginners, while adding more advanced exercises for more experienced participants. Both in the beginners' and advanced workshops, participants wished to cover topics in more detail, hinting towards the need for more instalments of the advanced workshops. Participants preferred structured exercises over free exercises. In the advanced workshop, they wished for further reading material and tools to study outside of the workshop and proposed to extend the workshop to a full day to cover more topics and details.

\subsection{Future outlook}
The education of users on how to sensibly and efficiently use available computing resources is key to reaching sustainable fundamental research. Therefore, the computing workshops will continue to teach these concepts to our institute members within the framework of the beginners' workshops. Furthermore, the single edition of the advanced workshop thus far had a positive reception, and, in addition to the feedback from the beginners' workshops, highlights the scope for more advanced workshops on various topics to be presented. 

The NAF is a national facility and, as such, has users from outside of DESY too. It would be great to expand the workshops to be able to include external users of the NAF as well, such that these tutorials on sustainable computing practices can reach as many users as possible. This extension of the workshops will be considered as the workshop series continues to grow and mature. Delivering the workshops in either a hybrid approach or a fully-virtual medium will be explored carefully to provide the optimum training with minimal impact from travel required.

Alongside the HSF training material that is already utilised in this workshop series, the authors of this paper hope that many similar initiatives at other HEP computing centres can evolve, perhaps using the structure developed here as a basis. If more such efforts are considered, a reduction in resources could be collectively achievable. Of course, more efficient use of resources does not necessarily lead to reduced usage, an effect known as the ``Rebound Effect''~\cite{sorrell2009rebound}; however, this is a topic that requires a much more thorough discussion that is outside the scope of what is presented in this paper. 

\section{Conclusions}
This perspective has presented the sustainable computing workshop series at DESY that aims to provide users of the NAF with the skills and knowledge to work in a more sustainable manner, hopefully leading to a reduced environmental impact by individuals. The series was launched in 2023 and has, to date, provided three beginners' workshops and one advanced workshop. Feedback from the workshops has shown that they have been positively received, and has resulted in a motivation to continue providing the workshops and expanding the program. Through continuous training and education, it will be possible to create a community of researchers that have a mindset that includes sustainability aspects when it comes to the use and employment of computing resources. 

\clearpage

\section*{Conflict of Interest Statement}
%All financial, commercial or other relationships that might be perceived by the academic community as representing a potential conflict of interest must be disclosed. If no such relationship exists, authors will be asked to confirm the following statement: 
The authors declare that the research was conducted in the absence of any commercial or financial relationships that could be construed as a potential conflict of interest.

\section*{Author Contributions}
The authors confirm that each author contributed to the production of this article.

\section*{Acknowledgements}
We acknowledge support from DESY (Hamburg, Germany), a member of the Helmholtz Association HGF, and support by the Deutsche Forschungsgemeinschaft (DFG, German Research Foundation) under Germany’s Excellence Strategy -- EXC 2121 ``Quantum Universe'' -- 390833306.

\bibliography{bibliography}

%%% Make sure to upload the bib file along with the tex file and PDF
%%% Please see the test.bib file for some examples of references

\end{document}